\documentclass[axioms,article,accept,moreauthors]{my-mdpi} 

\firstpage{1} 
\pubvolume{0}
\issuenum{0}
\articlenumber{0}
\pubyear{2021}
\copyrightyear{2021}

\datereceived{03 December 2020} 
\dateaccepted{25 January 2021} 
\datepublished{} 
\hreflink{} 

\usepackage{amsmath,amssymb,amsfonts,amsthm}
\usepackage{graphicx}
\usepackage{wrapfig}
\usepackage[labelfont=bf]{caption}
\usepackage{latexsym}
\usepackage[dvips]{epsfig}
\usepackage{url}
\usepackage{a4wide}

\allowdisplaybreaks

\pdfoutput=1

\Title{Modeling and Forecasting of COVID-19 Spreading
by Delayed Stochastic Differential Equations}

\TitleCitation{Modeling and Forecasting of COVID-19 Spreading
by Delayed Stochastic Differential Equations}

\Author{Marouane Mahrouf $^{1}$\orcidA{}, 
Adnane Boukhouima $^{1}$\orcidB{}, 
Houssine Zine $^{2}$\orcidC{}, 
El Mehdi Lotfi $^{1}$\orcidD{}, 
Delfim F. M. Torres $^{2,}$*\orcidE{} 
and Noura Yousfi $^{1}$\orcidF{}}

\AuthorNames{Marouane Mahrouf, Adnane Boukhouima, Houssine Zine, 
El Mehdi Lotfi, Delfim F. M. Torres and Noura Yousfi}

\AuthorCitation{Mahrouf,~M.; Boukhouima,~A.; Zine,~H.; Lotfi,~E.M.; Torres,~D.F.M.; Yousfi,~N.}

\address{$^{1}$ \quad Laboratory of Analysis, Modeling and Simulation (LAMS),
Faculty of Sciences Ben M'sik, Hassan II University of Casablanca,
Sidi Othman, P.B. 7955 Casablanca, Morocco; 
marouane.mahrouf@gmail.com~(M.M.);
adnaneboukhouima@gmail.com (A.B.); lotfiimehdi@gmail.com~(E.M.L.);
nourayousfi.fsb@gmail.com (N.Y.)\\
$^{2}$ \quad Center for Research and Development in Mathematics and Applications (CIDMA), 
Department of Mathematics, University of Aveiro, 3810-193 Aveiro, Portugal; zinehoussine@ua.pt}

\corres{Correspondence: delfim@ua.pt}

\abstract{The novel coronavirus disease (COVID-19) pneumonia has posed a great threat
to the world recent months by causing many deaths and enormous
economic damage worldwide. The first case of COVID-19 in Morocco
was reported on 2 March 2020, and the number of reported cases
has increased day by day. In this work, we extend the well-known SIR
compartmental model to deterministic and stochastic time-delayed
models in order to predict the epidemiological trend of COVID-19
in Morocco and to assess the potential role of multiple preventive
measures and strategies imposed by Moroccan authorities.
The main features of the work include the well-posedness
of the models and conditions under which the COVID-19 may become extinct
or persist in the population. Parameter values have been estimated from
real data and numerical simulations are presented
for forecasting the COVID-19 spreading as well
as verification of theoretical results.}


\begin{document}


\end{paracol}
\nointerlineskip


\section{Introduction}

Coronaviruses are a large family of viruses that cause illnesses,
ranging from the common cold to more serious illnesses such
as Middle Eastern Respiratory Syndrome (MERS-CoV) and
Severe Acute Respiratory Syndrome (SARS-CoV). The new coronavirus COVID-19
corresponds to a new strain that has not previously been identified in humans.
On 11 March 2020, COVID-19 was reclassified as a pandemic by the
World Health Organization (WHO). The disease has spread rapidly from country
to country, causing enormous economic damage and many deaths around the world,
prompting governments to issue a dramatic decree, ordering the lockdown 
of entire countries.

Since the confirmation of the first case of COVID-19 in Morocco
on 2 March 2020 in the city of Casablanca, numerous preventive
measures and strategies to control the spread of diseases have been
imposed by the Moroccan authorities. In addition, Morocco declared
a health emergency during the period from 20 March to 20 April 2020
and gradually extended it until 10 June 2020 in order to control the
spread of the disease. In this paper, we report the assessment of the evolution
of COVID-19 outbreak in Morocco. Besides shedding light on the dynamics
of the pandemic, the practical intent of our analysis is to provide officials
with the tendency of COVID-19 spreading, as well as gauge the effects
of preventives measures using mathematical tools.
Several other papers developed mathematical models for COVID-19 for particular 
regions in the globe and particular intervals of time, e.g., in~\cite{MR4164087}
a Susceptible--Infectious--Quarantined--Recovered (SIQR) model 
to the analysis of data from the Brazilian Department of Health, 
obtained from 26 February~2020 to 25~March 2020
is proposed to better understand the early evolution of COVID-19 in Brazil;
in~\cite{MR4128904}, a new COVID-19 epidemic model with media coverage 
and quarantine is constructed on the basis of the total confirmed new cases
in the UK from 1 February 2020 to 23 March 2020; while in \cite{MR4124320}
SEIR modelling to forecast the COVID-19 outbreak in Algeria is carried out 
by using available data from 1 March to 10 April, 2020. 

Mathematical modeling, particularly in terms of differential equations, 
is a strong tool that attracts the attention of many scientists to study various 
problems arising from mechanics, biology, physics, and so on. For instance, 
in \cite{Tanaka}, a system of differential equations with density-dependent 
sublinear sensitivity and logistic source is proposed and blow up properties
of solutions are investigated; paper \cite{Viglialoro} presents a mathematical 
model with application in civil engineering related to the equilibrium analysis 
of a membrane with rigid and cable boundaries; \cite{Pintus} studies
nonnegative and classical solutions to porous medium problems; and \cite{Li}
a two-dimensional boundary value problem under proper assumptions on the data. 
Herein, we will focus on the dynamic of COVID-19. \mbox{Tang et al. \cite{Tang}} 
used a Susceptible--Exposed--Infectious--Recovered (SEIR) compartmental model
to estimate the basic reproduction number of COVID-19 transmission, based
on data of confirmed cases for the disease in mainland China.
Wu et al. \cite{Wu} provided an estimate of the size of the epidemic in Wuhan
on the basis of the number of cases exported from Wuhan to cities outside
mainland China by using a SEIR model. In \cite{Kuniya}, Kuniya applied
the SEIR compartmental model for the prediction of the epidemic peak
for COVID-19 in Japan, using real-time data from 15 January to 29 February, 2020.
Fanelli and Piazza~\cite{Fanelli} analyzed and forecasted the COVID-19 spreading in China,
Italy and France, by using a simple Susceptible--Infected--Recovered--Deaths (SIRD) model.
A more elaborate model, which includes the transmissibility of super-spreader individuals,
is proposed in  \mbox{Nda\"irou et al. \cite{MR4093642}}. The model we propose here is new 
and has completely different compartments: in the paper \cite{MR4093642}, 
they model susceptible, exposed, symptomatic and infectious,
super-spreaders, infectious but asymptomatic, hospitalized, recovered and the fatality class, 
with the main contribution being the inclusion of super-spreader individuals; in contrast,
here we consider susceptible individuals, symptomatic infected individuals, which have not
yet been treated, the asymptomatic infected individuals who are infected but do not transmit
the disease, patients diagnosed and under quarantine and subdivided into three 
categories---benign, severe and critical forms---recovered and dead individuals. Moreover,
our model has delays, while the previous model \cite{MR4093642} has no delays;
our model is stochastic, while the previous model \cite{MR4093642} is deterministic. 
In fact, all mentioned models are deterministic and neglect the effect 
of stochastic noises derived from environmental fluctuations.
To the best of our knowledge, research works that predict the COVID-19 outbreak taking into account
a stochastic component, are a rarity \cite{S1,S2,S3}. The novelty of our work is twofold:
the extension of the models cited above to a more accurate model with time delay, suggested biologically
in the first place; secondly, to combine between the deterministic and the stochastic approaches
in order to well-describe reality. To do this, Section~\ref{sec:2} deals with the formulation
and the well-posedness of the models. Section~\ref{sec:3} is devoted to the qualitative
analysis of the proposed models. Parameters estimation and forecast of COVID-19
spreading in Morocco is presented in \mbox{Sections~\ref{sec:4} and \ref{sec:5}}, respectively.
The paper ends with discussion and conclusions, in Section~\ref{sec:6}.


\section{Models Formulation and Well-Posedness}
\label{sec:2}

Based on the epidemiological feature of COVID-19 and the several strategies
imposed by the government, with different degrees, to fight against this pandemic,
we extend the classical SIR model to describe the transmission of COVID-19
in the Kingdom of Morocco. In particular, we divide the population into eight
classes, denoted by $S$, $I_s$, $I_a$, $F_b$, $F_g$, $F_c$, $R$ and $M$, where $S$
represents the susceptible individuals; $I_s$ the symptomatic infected individuals,
which have not yet been treated; $I_a$ the asymptomatic infected individuals
who are infected but do not transmit the disease; $F_b$, $F_g$ and $F_c$ denote
the patients diagnosed, supported by the Moroccan health system and under quarantine,
and subdivided into three categories: benign, severe and critical forms, respectively.
Finally, $R$ and $M$ are the recovered and fatality classes. This model satisfies
the following assumptions:
\begin{itemize}
\item[(1)] all coefficients involved in the model are positive constants;
\item[(2)] natural birth and death rate are not factors;
\item[(3)] true asymptomatic patients will stay asymptomatic until recovery and do not spread the virus;
\item[(4)] patients who are temporarily asymptomatic are included on symptomatic ones;
\item[(5)] the second infection is not considered in the model;
\item[(6)] the Moroccan health system is not overwhelmed.
\end{itemize}

According to the above assumptions and the actual strategies imposed by the Moroccan authorities,
the spread of COVID-19 in the population is modeled by the following system
of delayed differential equations (DDEs):
\begin{equation}
\label{Sys1}
\left\{
\begin{array}{ll}
\dfrac{dS(t)}{dt}
&=-\beta(1-u)\dfrac{S(t)I_{s}(t)}{N},\\[0.3 cm]
\dfrac{dI_{s}(t)}{dt}
&=\beta\epsilon (1-u)\dfrac{S(t-\tau_1)I_{s}(t-\tau_1)}{N}-\alpha I_{s}(t)
-(1-\alpha)(\mu_s+\eta_{s})I_s(t),\\[0.3 cm]
\dfrac{dI_{a}(t)}{dt}
&=\beta(1-\epsilon)(1-u)\dfrac{S(t-\tau_1)I_{s}(t-\tau_1)}{N}-\eta_{a}I_{a}(t),\\[0.3 cm]
\dfrac{dF_{b}(t)}{dt}
&=\alpha\gamma_{b}I_{s}(t-\tau_2)-\big(\mu_b+r_b\big)F_{b}(t),\\[0.3 cm]
\dfrac{dF_{g}(t)}{dt}
&=\alpha\gamma_{g}I_{s}(t-\tau_2)-\big(\mu_g+r_g\big)F_{g}(t),\\[0.3 cm]
\dfrac{dF_{c}(t)}{dt}
&=\alpha\gamma_{c}I_{s}(t-\tau_2)-\big(\mu_c+r_c\big)F_{c}(t),\\[0.3 cm]
\dfrac{dR(t)}{dt}
&=\eta_{s}(1-\alpha)I_s(t-\tau_3)+\eta_{a}I_{a}(t-\tau_3)+r_{b}F_{b}(t-\tau_4)
+r_g F_{g}(t-\tau_4)+r_c F_{c}(t-\tau_4),\\[0.3 cm]
\dfrac{dM(t)}{dt}
&=\mu_s(1-\alpha)I_s(t-\tau_3)+\mu_bF_{b}(t-\tau_4)+\mu_gF_{g}(t-\tau_4)+\mu_cF_{c}(t-\tau_4),
\end{array}\right.
\end{equation}
where $t\in\mathbb{R}_{+}$, $N$ represents 
the total population size and $u\in[0,1]$ denotes the level
of the preventive strategies on the susceptible population.
The parameter $\beta$  indicates the transmission rate
and $\epsilon \in[0,1]$ is the proportion for the symptomatic individuals.
The parameter $\alpha$ denotes the proportion of the diagnosed symptomatic
infected population that moves to the three forms: $F_b$, $F_g$ and $F_c$,
by the rates $\gamma_b$, $\gamma_g$ and $\gamma_c$, respectively. The mean
recovery period of these forms are denoted by $1/r_b$, $1/r_g$ and $1/r_c$,
respectively. The latter forms die also with the rates $\mu_b$, $\mu_g$
and $\mu_c$, respectively. Asymptomatic infected population, which are not 
diagnosed, recover with rate $\eta_a$ and the symptomatic infected ones recover 
or die with rates $\eta_s$ and $\mu_s$, respectively. The time delays $\tau_1$, 
$\tau_2$, $\tau_3$ and $\tau_4$ denote the incubation period, the period of time 
needed before the charge by the health system, the time required before 
the death of individuals coming from the compartments $I_s$, $F_b$, $F_g$, 
and $F_c$, respectively. At each instant of time,
\begin{equation}
\mathcal{D}(t)=:\mu_s(1-\alpha)I_s(t-\tau_3)+\mu_bF_{b}(t-\tau_4)
+\mu_gF_{g}(t-\tau_4)+\mu_cF_{c}(t-\tau_4)=\dfrac{dM(t)}{dt}
\end{equation}
gives the number of new death due to the disease (cf. \cite{MR4093642}).

\begin{Remark}
In system \eqref{Sys1}, delays occur at the entrances, when the actions of infection 
take charge or the actions by the health system begin, and not at exits. 
Let us see an example. A susceptible individual, after contact with an infected person 
at instant $t$, becomes himself infected at instant $t+\tau_1$. 
Suddenly, the compartment of the infected is fed at the instant $t$ 
by the susceptible infected at the instant $t-\tau_1$. The same operation 
occurs at the level of the other interactions between the compartments of the model.
\end{Remark}

\begin{Remark}
We assume that the compartment of symptomatic infected $I_s$  
does not completely empty at any time $t$. For this reason,
one has $\mu_s+\eta_s<1$. Note also that the diagnosed 
symptomatic infected population is completely distributed into one
of three possible forms: $F_b$, $F_g$ and $F_c$, respectively
by the rates $\gamma_b$, $\gamma_g$ and $\gamma_c$. 
Then, $\gamma_b+\gamma_g+\gamma_c=1$.
\end{Remark}

\begin{Remark}
Biologically, $\tau_3 = 21$ days and $\tau_4 = 13.5$ days are the time periods 
needed before dying, deriving from $ I_s $ and the three forms $F_b,\ F_g,\ F_c$, 
respectively. That is why we inserted these delays in the last equation
of system \eqref{Sys1}.
\end{Remark}

\begin{Remark}
We consider only a short time period in comparison to the demographic time-frame.
From a biological point of view, this means that we can assume that there is neither 
entry (recruitment~rate) nor exit (natural mortality rate), and vital parameters 
can be neglected. Note also that in our model, the individuals that die due to the 
disease are included in the population. Therefore, the total population is here 
assumed to be constant, that is, $N(t) \equiv N$ during the period under study.
This assumption is also reinforced by the fact that the Moroccan authorities 
have closed geographic borders.
\end{Remark}

The initial conditions of system (\ref{Sys1}) are
\begin{equation}
\label{ic}
\begin{array}{ll}
S(\theta)
&=\varphi_{1}(\theta)\geq 0,
\quad I_s(\theta)=\varphi_{2}(\theta)\geq 0,
\quad I_a(\theta)=\varphi_{3}(\theta)\geq 0,\\
F_b(\theta)
&=\varphi_{4}(\theta)\geq 0,
\quad F_g(\theta)=\varphi_{5}(\theta)\geq 0,
\quad F_c(\theta)=\varphi_{6}(\theta)\geq 0,\\
R(\theta)
&=\varphi_{7}(\theta)\geq 0,
\quad M(\theta)=\varphi_{8}(\theta)\geq 0,
\quad \theta\in [-\tau,0],
\end{array}
\end{equation}
where $\tau=\max\{\tau_1,\tau_2,\tau_3,\tau_4\}$.
Let $\mathcal{C}=C([-\tau,0],\mathbb{R}^{8})$ be the Banach
space of continuous functions from the interval $[-\tau,0]$
into  $\mathbb{R}^{8}$ equipped with the uniform topology. It follows
from the theory of functional differential equations \cite{Hale} that
system (\ref{Sys1}) with initial conditions
$$
(\varphi_{1},\varphi_{2},\varphi_{3},\varphi_{4},\varphi_{5},
\varphi_{6},\varphi_{7},\varphi_{8})\in \mathcal{C}
$$
has a unique solution. On the other hand, due to continuous
fluctuation in the environment, the parameters of the system
are actually not absolute constants and always fluctuate randomly
around some average value. Hence, using delayed stochastic differential
equations (DSDEs) to model the epidemic provide some additional
degree of realism compared to their deterministic counterparts.
The parameters $\beta$ and $\alpha$ play an important role
in controlling and preventing COVID-19 spreading and they are not completely known,
but subject to some random environmental effects. We introduce randomness into system
(\ref{Sys1}) by applying the technique of parameter perturbation,
which has been used by many researchers (see, e.g.,
\cite{Mahrouf0,Mahrouf1,dalal}). In agreement, we replace the parameters
$\beta$ and $\alpha$ by $\beta \rightarrow \beta+\sigma_1 \dot{B}_1(t)$
and $\alpha \rightarrow \alpha+\sigma_2 \dot{B_2}(t)$, where
$B_{1}(t)$ and $B_{2}(t)$ are independent standard Brownian motions
defined on a complete probability space $(\Omega,\mathcal{F},\mathbb{P})$
with a filtration $\{\mathcal{F}_{t}\}_{t\geq0}$ satisfying the usual conditions
(i.e., it is increasing and right continuous while $\mathcal{F}_{0}$ contains
all P-null sets) and $\sigma_{i}$ represents the intensity of $B_{i}$ for $i=1,2$.
Therefore, we obtain the following model governed by delayed stochastic differential equations:
\begin{eqnarray}
\label{Sys2}
\fontsize{10}{10}{\left\{
\begin{array}{ll}
dS(t)
&=\left(-\beta(1-u)\dfrac{S(t)I_{s}(t)}{N}\right)dt
-\sigma_1(1-u)\dfrac{S(t)I_{s}(t)}{N}dB_1(t),\\[0.3 cm]
dI_{s}(t)
&=\left(\beta\epsilon (1-u)\dfrac{S(t-\tau_1)I_{s}(t-\tau_1)}{N}
-\alpha I_{s}(t) -(1-\alpha)(\mu_s+\eta_{s})I_s(t)\right)dt\\
&\quad +\sigma_1\left(\epsilon (1-u)\dfrac{S(t-\tau_1)I_{s}(t-\tau_1)}{N}\right)dB_1(t)
+\sigma_2(\mu_s+\eta_{s}-1)I_s(t)dB_2(t),\\[0.3 cm]
dI_{a}(t)
&=\left(\beta(1-\epsilon)(1-u)\dfrac{S(t-\tau_1)I_{s}(t-\tau_1)}{N}
-\eta_{a}I_{a}(t)\right)d(t)\\
&\quad +\sigma_1(1-\epsilon)(1-u)\dfrac{S(t-\tau_1)I_{s}
(t-\tau_1)}{N}dB_1(t),\\[0.3 cm]
dF_{b}(t)
&=\bigg(\alpha\gamma_{b}I_{s}(t-\tau_2)-\big(\mu_b+r_b\big)F_{b}(t)\bigg)dt
+\sigma_2\gamma_{b}I_{s}(t-\tau_2)dB_2(t),\\[0.3 cm]
dF_{g}(t)
&=\bigg(\alpha\gamma_{g}I_{s}(t-\tau_2)-\big(\mu_g+r_g\big)F_{g}(t)\bigg)dt
+\sigma_2\gamma_{g}I_{s}(t-\tau_2)dB_2(t),\\[0.3 cm]
dF_{c}(t)
&=\bigg(\alpha\gamma_{c}I_{s}(t-\tau_2)-\big(\mu_c+r_c\big)F_{c}(t)\bigg)dt
+\sigma_2\gamma_{c}I_{s}(t-\tau_2)dB_2(t),\\[0.3 cm]
dR(t)
&=\bigg(\eta_{s}(1-\alpha)I_s(t-\tau_3)+\eta_{a}I_{a}(t-\tau_3)
+r_{b}F_{b}(t-\tau_4)+r_g F_{g}(t-\tau_4)+r_c F_{c}(t-\tau_4)\bigg)dt\\
&\quad -\sigma_2\eta_{s}I_{s}(t-\tau_3)dB_2(t),\\[0.3 cm]
dM(t)
&=\left(\mu_s(1-\alpha)I_s(t-\tau_3)+\mu_bF_{b}(t-\tau_4)
+\mu_gF_{g}(t-\tau_4)+\mu_cF_{c}(t-\tau_4)\right)dt\\
&\quad -\sigma_2\mu_{s}I_{s}(t-\tau_3)dB_2(t),
\end{array}\right.}
\end{eqnarray}
where the coefficients are locally Lipshitz 
with respect to all the variables, for all $ t\in \mathcal{R}^{+}$.

Let us denote
$\mathbb{R}^8_+=\{(x_1,x_2,x_3,x_4,x_5,x_6,x_7,x_8)\mid x_i>0,
\ i=1,2,\ldots,8\}$. We have the following result.

\begin{Theorem}
For any initial value satisfying condition \eqref{ic},
there is a unique solution
$$
x(t)=(S(t),I_s(t),I_a(t),F_b(t),F_g(t),F_c(t),R(t),M(t))
$$
to the COVID-19 stochastic model \eqref{Sys2} that remains
in $\mathbb{R}^8_+$ with a probability of one.
\end{Theorem}

\begin{proof}
Since the coefficients of the stochastic differential equations
with several delays~\eqref{Sys2} are locally Lipschitz continuous,
it follows from \cite{Mao} that for any square integrable initial
value $x(0)\in\mathbb{R}^{8}_{+}$, which is independent of the
considered standard Brownian motion $B$, there exists a unique
local solution $x(t)$ on $t\in[0,\tau_e)$, where  $\tau_e$ is the explosion time.
For showing that this solution is global, knowing that the linear growth
condition is not verified, we need to prove that $\tau_e=\infty$.
Let $k_0>0$ be sufficiently large for $\dfrac{1}{k_0}<x(0)<k_0$.
For each integer $k\geq k_0$, we define the stopping time
$\tau_k = \inf\left\lbrace t\in[0,\tau_e)\text{ s.t. } x_i(t)\notin
\left( \dfrac{1}{k},k\right)\; \text{for some}\;
i=1,2,3\right\rbrace$, where $\inf \emptyset=\infty$.
It is clear that $\tau_k\leq \tau_e$. Let $T>0$ be arbitrary. Define the
twice differentiable function $W$ on $\mathbb{R}^{{*}^{3}}_{+}\rightarrow \mathbb{R}^+$
as follows:
$$
W(x)=(x_1+x_2+x_3)^2+\frac{1}{x_1}+\frac{1}{x_2}+\frac{1}{x_3}.
$$
By It\^o's formula, for any $0\leq t\leq \tau_k \wedge T$
and $k\geq 1$, we have
$$
dW(x(t))=LW(x(t))dt+\zeta(x(t))dB(t),
$$
where $\zeta$ is a continuous functional defined on 
$[0,+\infty)\times C([-\tau,0],\mathbb{R}^{3\times2})$ by 
\begin{equation*}
\zeta(x(t))=\left(\begin{array}{cc}
-\sigma_1(1-u)\dfrac{S(t)I_{s}(t)}{N} & 0 \\
\sigma_1\epsilon (1-u)\dfrac{S(t-\tau_1)I_{s}(t-\tau_1)}{N} 
& \sigma_2(\mu_s+\eta_{s}-1)I_s(t) \\
\sigma_1(1-\epsilon)(1-u)\dfrac{S(t-\tau_1)I_{s}
(t-\tau_1)}{N} & 0
\end{array}\right),
\end{equation*}
$B(t)=(B_1(t),B_2(t))^{\mathcal{T}}$ with the superscript ``$\mathcal{T}$'' 
representing transposition, and $L$ is the differential operator 
of function $W$ defined by
\begin{equation*}
\begin{split}
L&W(x(t))
= \left(2(S(t)+I_s(t)+I_a(t))-\dfrac{1}{S^2(t)}\right)\left(
-\beta (1-u)\dfrac{S(t)I_s(t)}{N}\right) \\
&+ \left(1+\dfrac{1}{S^3(t)}\right)\left(-\sigma_1(1-u)
\dfrac{S(t) I_s(t)}{N}\right)^2 \\
&+ \left(2(S(t)+I_s(t)+I_a(t))-\dfrac{1}{I_{s}^{2}(t)}\right)
\left[\beta\epsilon(1-u)\dfrac{S(t-\tau_1)I_s(t-\tau_1)}{N}
-\alpha I_s(t)-(1-\alpha)(\mu_s+\eta_s)I_s(t)\right] \\
&+ \left( 1+\frac{1}{I_{s}^3(t)}\right) \left[
\left( \sigma_1\epsilon(1-u)\dfrac{S(t-\tau_1)
I_s(t-\tau_1)}{N}\right)^2
+\big(\sigma_2(\mu_s+\eta_s-1)I_s(t) \big)^2\right]\\
&+ \left( 2(S(t)+I_s(t)+I_a(t))-\frac{1}{I^2_a(t)}\right)
\left(\beta(1-\epsilon)(1-u)\dfrac{S(t-\tau_1)
I_s(t-\tau_1)}{N}-\eta_aI_a(t)\right)\\
&+\left( 1+\dfrac{1}{I_{a}^3(t)}\right)\left(
\sigma_1(1-\epsilon)(1-u)\dfrac{S(t-\tau_1)I_s(t-\tau_1)}{N}\right)^2.
\end{split}
\end{equation*}
Thus,
\begin{equation}
\label{eq:ineq}
\begin{split}
LW(x(t))
&\leq  \dfrac{\beta (1-u)S(t) I_s(t)}{NS^{2}(t)}
+\left(1+\dfrac{1}{S^3(t)}\right)\left(\sigma_1(1-u)\dfrac{S(t) I_s(t)}{N}\right)^2\\
&+ 2\beta\epsilon(1-u)\big(S(t)+I_s(t)+I_a(t)\big)\dfrac{S(t-\tau_1)
I_s(t-\tau_1)}{N}+\dfrac{\alpha+(1-\alpha)(\mu_s+\eta_s)}{I_{s}(t)} \\
&+ \left(1+\dfrac{1}{I_{s}^{3}(t)}\right)\left[
\left(\sigma_1\epsilon(1-u)\dfrac{S(t-\tau_1) I_s(t-\tau_1 )}{N}\right)^2
+\big(\sigma_2(\mu_s+\eta_s-1)I_s(t) \big)^2\right]\\
&+ 2\beta(1-\epsilon)(1-u)\big(S(t)+I_s(t)+I_a(t)\big)
\dfrac{S(t-\tau_1) I_s(t-\tau_1)}{N}\\
&+ \dfrac{\eta_a}{I_a(t)}+\left(1+\dfrac{1}{I_{a}^{3}(t)}\right)
\left(\sigma_1(1-\epsilon)(1-u)\dfrac{S(t-\tau_1) I_s(t-\tau_1)}{N}\right)^2.
\end{split}
\end{equation}
We now apply the elementary inequality $2xy\leq x^2+y^2$, 
valid for any $x,y \in \mathbb{R}$, by firstly taking
$x=\beta\epsilon(1-u)$ and $y=S(t)+I_s(t)+I_a(t)$ and, secondly, 
$x=\beta(1-\epsilon)(1-u)$ and $y=S(t)+I_s(t)+I_a(t)$. In this way, we easily
increase the right-hand side of inequality \eqref{eq:ineq} to obtain that
\begin{eqnarray*}
LW(x(t))&\leq& b_1+\psi\big(S(t)+I_s(t)+I_a(t)\big)^2
+\dfrac{b_2}{S(t)}+\dfrac{b_3}{I_s(t)}+\dfrac{b_4}{I_a(t)}\\
&\leq& D(1+W(x(t))),
\end{eqnarray*}
where $\psi,\ b_1,\ b_2,\ b_3, \text{ and } b_4$ are positive constants
and $D=\max\left(\psi,\ b_1,\ b_2,\ b_3,\ b_4\right)$.
By integrating both sides of equality
$$
dW(x(t))=LW(x(t))dt+\zeta(x(t))dB(t)
$$
between $t_0$ and $t\wedge \tau_k$ and acting the expectation,
which eliminates the martingale part, we~get 
\begin{eqnarray*}
E(W(x(t\wedge \tau_k))
&=& E(W(x_0))+E\int^{t\wedge \tau_k}_{t_0}LW(x(s)))ds\\
&\leq & E(W(x_0))+E\int^{t\wedge \tau_k}_{t_0}D(1+W(x(s)))ds\\
&\leq &  E(W(x_0))+DT+\int^{t \wedge \tau_k}_{t_0}EW(x(s)))ds
\end{eqnarray*}
and Gronwall's inequality implies that
$$
E(W(x(t\wedge \tau_k))\leq (EW(x_0)+DT)\exp(CT).
$$
For $\omega \in \{\tau_k\leq T\}$, $x_i(\tau_k)$
equals $k$ or $\dfrac{1}{k}$ for some $ i=1,2,3$.
Hence,
$$
W(x_i(\tau_k))\geq \left(k^2+\dfrac{1}{k}\right)
\wedge \left(\dfrac{1}{k^2}+k\right).
$$
It follows that
\begin{eqnarray*}
(EW(x_0)+DT)\exp(CT)
&\geq & E\left( \chi_{\{\tau_k \leq T\}}(\omega) W(x_{\tau_k})\right) \\
&\geq & \left(k^2+\dfrac{1}{k}\right)
\wedge \left(\dfrac{1}{k^2}+k\right)
P(\tau_k \leq T).
\end{eqnarray*}

Letting $k\rightarrow \infty$, we get $P(\tau_e \leq T)=0$.
Since $T$ is arbitrary, we obtain $P(\tau_e =\infty)=1$.
\textls[-15]{By defining the stopping time} 
$\tilde{\tau}_k = \inf\left\lbrace t\in[0,\tau_e)\text{ s.t. } 
x_i(t)\notin \left( \dfrac{1}{k},k\right)\; \text{for some}\;
i=4,\ldots,8\right\rbrace$, and considering the twice differentiable function 
$\tilde{W}$ on $\mathbb{R}^{{*}^{5}}_{+}\rightarrow \mathbb{R}^+$ as
$$
\tilde{W}(x)=\left( \sum_{i=4}^{8}x_i\right) ^2+\sum_{i=4}^{8}\frac{1}{x_i},
$$
we deduce, with the same technique, that all the variables 
of the system are positive on $[0,\infty)$.
\end{proof}


\section{Qualitative Analysis of the Models}
\label{sec:3}

The basic reproduction number, as a measure for disease spread
in a population, plays an important role in the course and control
of an ongoing outbreak \cite{Diekmann}. This number is defined as the
expected number of secondary cases produced, in a completely susceptible population,
by a typical infective individual. Note that the calculation of the basic reproduction number 
$R_0$ does not depend on the variables of the system but depends on its parameters.
In addition, the $R_0$ of our model does not depend on the time delays.
For this reason, we use the next-generation matrix approach outlined
in \cite{van:den:Driessche} to compute $R_0$. Precisely, 
the basic reproduction number $\mathcal{R}_{0}$
of system (\ref{Sys1}) is given by
\begin{equation}
\label{18}
\mathcal{R}_{0}=\rho(FV^{-1})=\dfrac{\beta\epsilon(1-u)}{(1-\alpha)(\eta_{s}+\mu_{s})+\alpha},
\end{equation}
where $\rho$ is the spectral radius of the next-generation matrix $ FV^{-1} $ with
\begin{center}
$F=\begin{pmatrix}
\beta \epsilon (1-u) & 0 \\
0 & 0
\end{pmatrix}$
\quad and \quad
$ V= \begin{pmatrix}
(1-\alpha)(\eta_{s}+\mu_{s})+\alpha& 0 \\
0 & \eta_{a}
\end{pmatrix}$.
\end{center}

Noting that the classes that are directly involved in the spread of disease are only
$ I_s $, $ I_a $, $ F_b $, $ F_g $ and $ F_c $, we can reduce the local stability
of system \eqref{Sys1} to the local stability of
\begin{equation}
\label{Sys6}
\fontsize{10}{10}{\left\{
\begin{array}{ll}
\dfrac{dI_{s}(t)}{dt}
&=\beta\epsilon (1-u)\dfrac{S(t-\tau_1)I_{s}(t-\tau_1)}{N}
-\alpha I_{s}(t)-(1-\alpha)(\mu_s+\eta_{s})I_s(t),\\[0.3 cm]
\dfrac{dI_{a}(t)}{dt}
&=\beta(1-\epsilon)(1-u)
\dfrac{S(t-\tau_1)I_{s}(t-\tau_1)}{N}-\eta_{a}I_{a}(t),\\[0.3 cm]
\dfrac{dF_{b}(t)}{dt}
&=\alpha\gamma_{b}I_{s}(t-\tau_2)
-\big(\mu_b+r_b\big)F_{b}(t),\\[0.3 cm]
\dfrac{dF_{g}(t)}{dt}
&=\alpha\gamma_{g}I_{s}(t-\tau_2)-\big(\mu_g+r_g\big)F_{g}(t),\\[0.3 cm]
\dfrac{dF_{c}(t)}{dt}
&=\alpha\gamma_{c}I_{s}(t-\tau_2)-\big(\mu_c+r_c\big)F_{c}(t).
\end{array}\right.}
\end{equation}
The other classes are uncoupled to the equations of system \eqref{Sys1}
and the total population size $N$ is constant. Then, we can easily
obtain the following analytical results:
\begin{equation}
\fontsize{10}{10}{\left\{
\begin{split}
S(t)& = N-\big( I_{s}(t)+I_{a}(t)+F_{b}(t)+ F_{g}(t)+ F_{c}(t)+ R(t) + M(t)\big),\\
R(t)& = \int^{t}_{0}\left[\eta_{s}(1-\alpha)I_s(\delta-\tau_3)
+\eta_{a}I_{a}(\delta-\tau_3)+r_{b}F_{b}(\delta-\tau_4)
+r_g F_{g}(\delta-\tau_4)+r_c F_{c}(\delta-\tau_4)\right] d\delta,\\[0.3 cm]
M(t)&=\int^{t}_{0}\left[\mu_s(1-\alpha)I_s(\delta-\tau_3)
+\mu_aI_a(\delta-\tau_3)+\mu_bF_{b}(\delta-\tau_4)
+\mu_gF_{g}(\delta-\tau_4)+\mu_cF_{c}(\delta-\tau_4)\right]d\delta.
\end{split}\right.}
\end{equation}

Let $\overline{E}=(\overline{I_s},\overline{I_a},\overline{F_b},
\overline{F_g},\overline{F_c})$ be an arbitrary equilibrium, 
and consider into system (\ref{Sys6}), the following change of unknowns:
$$
U_1(t)=I_s(t)-\overline{I_s},\,
U_2(t)=I_a(t)-\overline{I_a},\,
U_3(t)=F_b(t)-\overline{F_b},
U_4(t)=F_g(t)-\overline{F_g} 
\text{ and } 
U_5(t)=F_c(t)-\overline{F_c}.
$$
By substituting $U_i(t)$, $i=1,2,\ldots,5$, into system (\ref{Sys6}) 
and linearizing around the free equilibrium, we get a new system 
that is equivalent to
\begin{equation}
\dfrac{dX(t)}{dt}= A X(t) + B X(t-\tau_1) + C X(t-\tau_2),
\end{equation}
where $X(t)=(U_1(t),U_2(t),U_3(t),U_4(t),U_5(t))^\mathcal{T}$ 
and  $ A $, $ B $, $ C $ are the Jacobian
matrix \mbox{of (\ref{Sys6})} given by
\begin{equation*}
\begin{array}{ccccc}
A&=&\left(\begin{array}{ccccc}
-\alpha-(1-\alpha)(\mu_s+\eta_s) & 0 & 0 &0 & 0\\
0 & -\eta_a & 0 &0 &0 \\
0& 0 & -(\mu_b+r_b) & 0&0 \\
0& 0 & 0 &-(\mu_g+r_g) & 0 \\
0 & 0 & 0 &0 & -(\mu_c+r_c)
\end{array}\right),
\end{array}
\end{equation*}
\begin{equation*}
\begin{array}{ccccc}
B& =&\left(\begin{array}{ccccc}
\beta \epsilon (1-u) & 0 & 0 &0 & 0\\
\beta (1-\epsilon) (1-u)&0 & 0 &0 &0 \\
0& 0 & 0 & 0&0 \\
0& 0 & 0 &0 & 0 \\
0 & 0 & 0 &0 & 0
\end{array}\right),
\end{array}
\end{equation*}
and
\begin{equation*}\label{N}
\begin{array}{ccccc}
C& =&\left(\begin{array}{ccccc}
0 & 0 & 0 &0 & 0\\
0&0 & 0 &0 &0 \\
\alpha \gamma_b& 0 & 0 & 0&0 \\
\alpha \gamma_g& 0 & 0 &0 & 0 \\
\alpha \gamma_c& 0 & 0 &0 & 0
\end{array}\right).
\end{array}
\end{equation*}

The characteristic equation of system (\ref{Sys6}) is given by
\begin{equation}
\label{EC}
P(\lambda)=(\lambda-a_1(\mathcal{R}_{0}e^{-\lambda\tau_1}-1))(
\lambda+\eta_a)(\lambda+(\mu_b+r_b))(\lambda+(\mu_g+r_g))(\lambda+(\mu_c+r_c)),
\end{equation}
where
\begin{equation*}
\begin{array}{ll}
a_1= & \alpha+(1-\alpha)(\mu_s+\eta_s).
\end{array}
\end{equation*}
Clearly, the characteristic Equation \eqref{EC} has the roots
$\lambda_{1}=- \eta_a$,  $\lambda_{2}=- (\mu_b+r_b) $,
$ \lambda_{3}=- (\mu_g+r_g) $, $ \lambda_{4}=- (\mu_c+r_c) $
and the root of the equation
\begin{equation}
\label{SC}
\lambda-a_1(\mathcal{R}_{0}e^{-\lambda \tau_1}-1)=0.
\end{equation}

We suppose $ Re(\lambda)\geq 0 $. From \eqref{SC}, we get
\begin{equation*}
Re(\lambda)= a_1(\mathcal{R}_{0}e^{-Re(\lambda) \tau_1}\cos(Im\lambda\; \tau_1)-1)<0,
\end{equation*}
if $ \mathcal{R}_{0}<1 $, which contradicts
$Re(\lambda)\geq 0$. On the other hand, we show that \eqref{SC} has a real
positive root when $\mathcal{R}_{0}>1 $. Indeed, we put
\begin{equation*}
\Phi(\lambda)= \lambda-a_1(\mathcal{R}_{0}e^{-\lambda \tau_1}-1).
\end{equation*}

We have that $\Phi(0)=- a_1(\mathcal{R}_{0}-1) <0 $,
$\lim_{\lambda\to +\infty}\Phi(\lambda)=+\infty$ and function
$ \Phi  $ is continuous on $ (0,+\infty) $. Consequently,
$\Phi$ has a positive root and the following result holds.

\begin{Theorem}
\label{thm:3.1}
The disease free equilibrium of system (\ref{Sys1}), that is,
$(N,0,0,0,0,0,0,0)$, is locally asymptotically stable if $\mathcal{R}_{0}< 1$
and unstable if $\mathcal{R}_{0}>1$.
\end{Theorem}

Knowing the value of the deterministic threshold $\mathcal{R}_{0}$
characterizes the dynamical behavior of system (\ref{Sys1}) and
guarantees persistence or extinction of the disease. Similarly,
now we characterize the dynamical behavior of system (\ref{Sys2})
by a sufficient condition for extinction of the disease.

\begin{Theorem}
\label{thm:3.2}
Let $x(t)=\big(S(t),I_s(t),I_a(t),F_b(t),F_g(t),F_c(t),R(t),M(t)\big)$
be the solution of the COVID-19 stochastic model (\ref{Sys2})
with initial value $x(0)$ defined in (\ref{ic}). Assume that
$$
\sigma_{1}^2>\dfrac{\beta^2}{2(\alpha+(1-\alpha)(\mu_s+\eta_s))}.
$$
Then,
\begin{equation}
\label{th1}
\limsup_{t\rightarrow+\infty}\ln\dfrac{I_s(t)}{t}<0.
\end{equation}
Namely, $I_s(t)$ tends to zero exponentially almost surely,
that is, the disease dies out with a probability of one.
\end{Theorem}

\begin{proof}
Let
\begin{align*}
d\ln I_s(t)
=& \left[ \dfrac{1}{I_s(t)}\left( \beta \epsilon (1-u)
\dfrac{S(t-\tau_1)I_s(t-\tau_1)}{N} - \alpha I_s(t)
- (1-\alpha)(\mu_s+\eta_s)I_s(t)\right) \right. \\
& \left. -\dfrac{1}{2I_s^{2}(t)}\left( \left(\sigma_1 \dfrac{\beta
\epsilon (1-u)S(t-\tau_1)I_s(t-\tau_1)}{N} \right)^{2}
+ \big( \sigma_2(\mu_s+\eta_s-1)I_s(t)\big)^{2}\right)\right] dt\\
&+\sigma_1\beta \epsilon (1-u) \dfrac{S(t-\tau_1)I_s(t-\tau_1)}{NI_s(t)} dB_1(t)
+ \sigma_2 (\mu_s+\eta_s-1) dB_{2}(t).
\end{align*}
To simplify, we set
\begin{gather*}
G(t)=\epsilon(1-u)\dfrac{S(t-\tau_1) I_s(t-\tau_1)}{N},
\quad R_{1}(t)= \sigma_1\beta\dfrac{ G(t)}{I_s(t)},\\
R_{3} = \sigma_2 (\mu_s+\eta_s-1), 
\quad H = -\alpha-(1-\alpha)(\mu_s+\eta_s).
\end{gather*}
Then, we get
\begin{align*}
d\ln I_s(t)
=& \,  \left[\dfrac{\beta G(t)}{I_s(t)} + H -\dfrac{1}{2}
\left( \left( \dfrac{\sigma_1 G(t)}{I_s(t)} \right)^{2}
+ R_{3}^{2}\right)\right]dt+ R_1(t) dB_{1}(t) + R_3 dB_{2}(t)\\
=& \left[-\dfrac{\sigma_1^{2}}{2}\left[ \left(
\dfrac{G(t)}{I_s(t)}\right) ^{2}- \dfrac{2\beta}{\sigma_1^{2}}
\dfrac{G(t)}{I_s(t)}\right] + H 
-\dfrac{R_{3}^{2}}{2} \right]dt+ R_1(t) dB_{1}(t)+ R_3dB_{2}(t)\\
=& \left[-\dfrac{\sigma_1^{2}}{2} \left[ \left( \dfrac{G(t)}{I_s(t)}
- \dfrac{\beta}{\sigma_1^{2}} \right) ^{2} - \dfrac{\beta^{2}}{
\sigma_1^4} \right] + H -\dfrac{R_{3}^{2}}{2} \right]dt+ R_1(t) dB_{1}(t) + R_3 dB_{2}(t)\\
\leq& \,\left[\dfrac{\beta^{2}}{2\sigma_1^{2}}
+ H \right]dt+ R_1(t) dB_{1}(t) + R_3dB_{2}(t).
\end{align*}
Integrating both sides of the above inequality between $0$ and $t$,  one has
\begin{equation*}
\dfrac{\ln I_s(t)}{t}
\leq \, \dfrac{\ln I_s(0)}{t}
+ \dfrac{\beta^{2}}{2\sigma_1^{2}}
+ H + \dfrac{M_1(t)}{t} + \dfrac{M_3(t)}{t} ,
\end{equation*}
where
\begin{equation*}
M_1(t)= \int_{0}^{t}R_{1}(s)dB_{1}(s)
\quad \text{and}\quad
M_3(t)= \int_{0}^{t}R_{3}dB_{2}(s).
\end{equation*}
We have
\begin{eqnarray*}
<M_1,M_1>_t
&=&  \int^t_0{\sigma_1}^2\epsilon^2(1-u)^2
\dfrac{S(s-\tau_1)^2I_s(s-\tau_1)^2}{N^2I_s^2(s)}ds\\
&\leq &  \int^t_0{\sigma_1}^2\epsilon^2(1-u)^2
\dfrac{N^4}{N^2}\dfrac{1}{I^2_s(s)}ds\\
&\leq &  \int^t_0{\sigma_1}^2\epsilon^2(1-u)^2ds.
\end{eqnarray*}
Then,
$$
\underset{t\rightarrow\infty}{\limsup}\dfrac{<M_1,M_1>_t}{t}
\leq {\sigma_1}^2\epsilon^2(1-u)^2<+\infty.
$$
From the large number theorem for martingales \cite{Grai}, we deduce that
$$
\underset{t\rightarrow\infty}{\lim}\dfrac{M_1(t)}{t}=0.
$$
We also have
\begin{equation*}
<M_3,M_3>_t
= \int^t_0 \sigma^2_3(\mu_s+\eta_s-1)^2ds
= \sigma^2_3(\mu_s+\eta_s-1)^2t.
\end{equation*}
Then,
$$
\underset{t\rightarrow\infty}{\limsup}\dfrac{<M_3,M_3>_t}{t}
\leq \sigma^2_3(\mu_s+\eta_s-1)<+\infty
$$
and
$$
\underset{t\rightarrow\infty}{\lim}\dfrac{M_3(t)}{t}=0.
$$
Subsequently,
$$
\limsup_{t\rightarrow+\infty}\ln\dfrac{I_s(t)}{t}
\leq \dfrac{\beta^2}{2\sigma_1^2}-\alpha-(1-\alpha)(\mu_s+\eta_s).
$$
We conclude that if $\dfrac{\beta^2}{2\sigma_1^2}-\alpha-(1-\alpha)(\mu_s+\eta_s)<0$,
then $\underset{t\rightarrow\infty}{\lim I(t)}=0$. This completes the proof.
\end{proof}


\section{Assessment of Parameters}
\label{sec:4}

Estimating the model parameters poses a big challenge because the COVID-19
situation changes rapidly and from one country to another. The parameters are
likely to vary over time as new policies are introduced on a day-to-day basis.
For this reason, in order to simulate the COVID-19 models \eqref{Sys1} and \eqref{Sys2},
we consider some parameter values from the literature, while the remaining
ones are estimated or fitted.

As the transmission rate $\beta$ is unknown, we carry out the least-square method \cite{Kuniya}
to estimate this parameter, based on the actual official reported confirmed cases
from 2~March to 20 March, 2020 \cite{url:HIVdata:morocco}. Through this method,
we estimated $ \beta $ as $0.4517$ (95\%CI, 0.4484--0.455). Since the life expectancy
for symptomatic individuals is 21 days on average and the crude mortality ratio
is between $3\%$ to $4\%$ \cite{WHO}, we estimated $\mu_s=0.01/21 $ per day and
$ \eta_s= 0.8/21$ per day. Furthermore, since the hospitals are not yet saturated
and the epidemic situation is under control, we assume that mortality comes
mainly from critical forms with a percentage of $40\%$ for an average period
of 13.5 days \cite{WHO}. Then, we choose $ \mu_c= 0.4/13.5 $ per day and
$r_c = 0.6/13.5 $ per day. According to \cite{Mizumoto}, the proportion
of asymptomatic individuals varies from $20.6 \%$ to $39.9 \%$ and of symptomatic
individuals from $60.1 \%$ and $79.4 \%$ of the infected population.
The progression rates $ \gamma_b $, $ \gamma_g $ and $ \gamma_c $,
from symptomatic infected individuals to the three forms,
are assumed to be $80 \%$ of diagnosed cases for benign form,
$15\%$ of diagnosed cases for severe form, and $5\%$ of diagnosed cases
for critical form, respectively \cite{WHO}. The incubation period
is estimated to be $5.5$ days \cite{WHO1,Stephen} while the time needed
before hospitalization is to be $7.5$ days \cite{Huang,Wang,Haut}.
Following a clinical observation related to the situation of COVID-19 in Morocco,
an evolution of symptomatic individuals is estimated towards recovery or death
after 21 days without any clinical intervention. In the case when
clinical intervention is applied, we estimate the evolution of the critical
forms towards recovery or death after 13.3 days.
The rest of the parameter values are shown in Table~\ref{values1}.
\begin{specialtable}[H]
\centering	
\tablesize{\small}
\captionof{table}{Parameter values of models (\ref{Sys1}) and (\ref{Sys2}).}
\label{values1}
\setlength{\tabcolsep}{3.8mm}
\begin{tabular}{cccccc}
\toprule
\textbf{Parameter} &   \textbf{Value} & \textbf{Source} & \textbf{Parameter} &   \textbf{Value} & \textbf{Source} \\
\midrule
$\beta$ & $0.4517 $& Estimated& $u$ & [0--1]  & Varied \\
$\epsilon$ &   $0.794$ & \cite{Mizumoto} & $\gamma_{b}$ &  $0.8$ & \cite{WHO}\\
$\gamma_{g}$ &  $0.15$ & \cite{WHO}& $\gamma_{c}$ & $0.05$ & \cite{WHO} \\
$\alpha$ &  0.06 & Assumed & $\eta_{a}$ & $1/21$& Calculated\\
$\eta_{s}$ &  $0.8/21$& Calculated & $\mu_{s}$ &   $0.01/21$& Calculated\\
$\mu_{b}$ &  $0$& Assumed & $\mu_{g}$ &  $0$& Assumed\\
$\mu_{c}$ &  $0.4/13.5$& Calculated & $r_{b}$ & $ 1/13.5$& Calculated\\
$r_{g}$ & $1/13.5$& Calculated & $r_{c}$ &  $0.6/13.5$& Calculated\\
$ \tau_1 $& $ 5.5$& \cite{WHO1,Stephen}& $ \tau_2 $&  $7.5$& \cite{Huang,Wang,Haut}\\
$ \tau_3 $& $ 21 $& Assumed & $ \tau_4 $&  $13.5$& Assumed\\
$ \sigma_1 $& $ 1.03 $& Calculated & $ \sigma_2 $&  $0.1$& Assumed\\
\bottomrule
\end{tabular}
\end{specialtable}


\section{Numerical Simulation of Moroccan COVID-19 Evolution}
\label{sec:5}

In this section, we present the forecasts of COVID-19 in Morocco related
to different strategies implemented by Moroccan authorities.

\begin{figure}[H]
\centering	
\includegraphics[width=15cm,height=7cm]{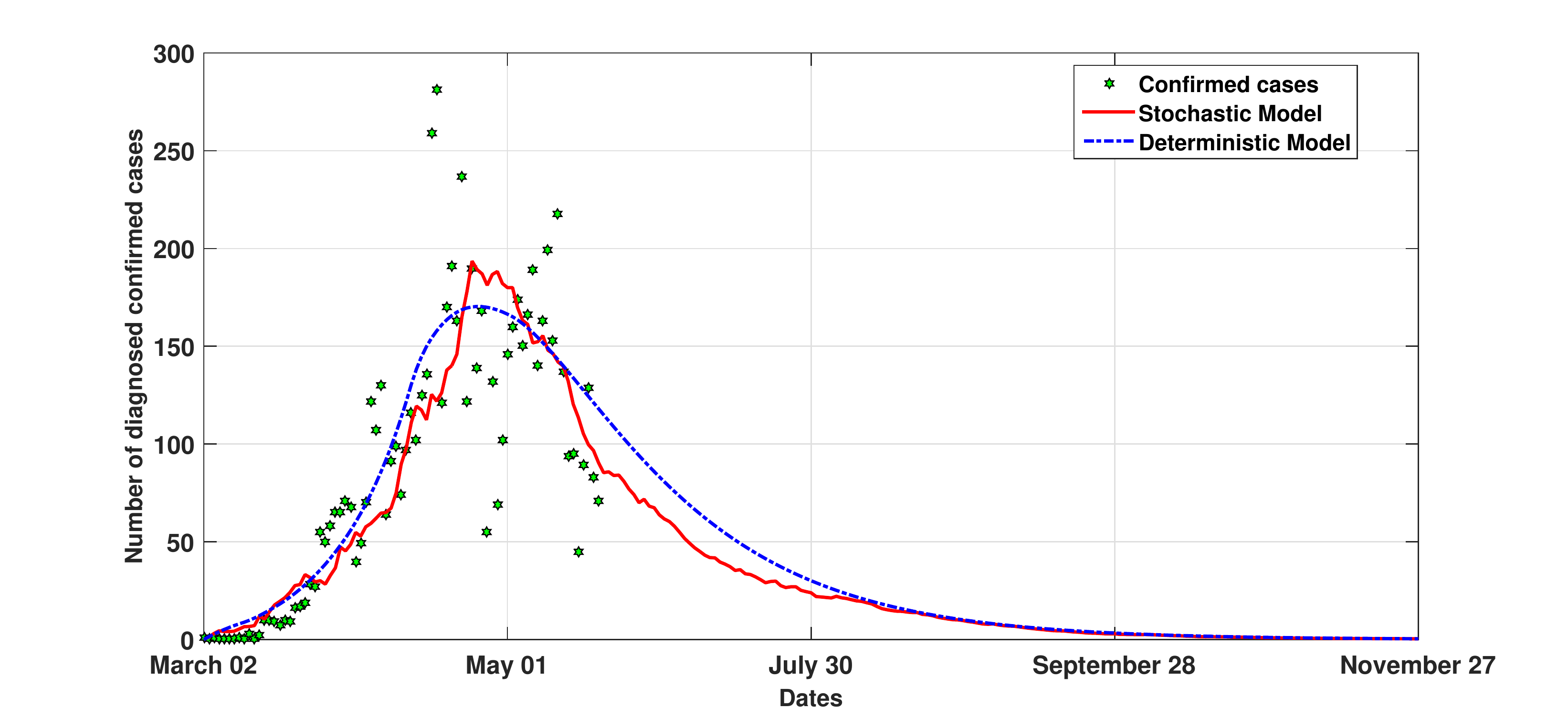}
\captionof{figure}{Comparison of the deterministic and the stochastic
dynamical behavior with the daily reported cases of COVID-19 in Morocco.}
\label{Historique}
\end{figure}

Taking into account the four levels of measures attached
to containment, the effectiveness level of the applied
Moroccan preventive measures is estimated to be
$$
u=\left\{
\begin{array}{ll}
0.2, & \hbox{on $ (\text{2 March},\text{10 March]}$;} \\
0.3, & \hbox{on $(\text{10 March},\text{20 March]}$;}\\
0.4, & \hbox{on $(\text{20 March},\text{6 April]}$;}\\
0.8, & \hbox{after $ \text{6 April}$.}
\end{array}
\right.
$$
In Figure~\ref{Historique}, we see that the plots and the clinical
data are globally homogeneous.
In addition, the last daily reported cases in Morocco \cite{4},
confirm the biological tendency of our model.
Thus, our models are efficient to describe the spread of \mbox{COVID-19} in Morocco.
However, we note that some clinical data are far from the values
of the models due to certain foci that appeared in some large areas or at the level
of certain industrial areas. We conclude also that the stochastic behavior
of COVID-19 presents certain particularities contrary to the deterministic one,
namely the magnitude of its peak is higher and the convergence to eradication is faster.
On the other hand, the conditions in Theorems~\ref{thm:3.1} and \ref{thm:3.2}
are verified. More precisely, the basic reproduction number $\mathcal{R}_{0}=0.5230$
is less than one from 12 May 2020 and $\sigma_{1}^{2}=1.0609>1.0598
=\dfrac{\beta^2}{2(\alpha+(1-\alpha)(\mu_s+\eta_s))}$,
which means that the eradication of disease is ensured.

\begin{figure}[H]
\centering
\includegraphics[width=15cm,height=7cm]{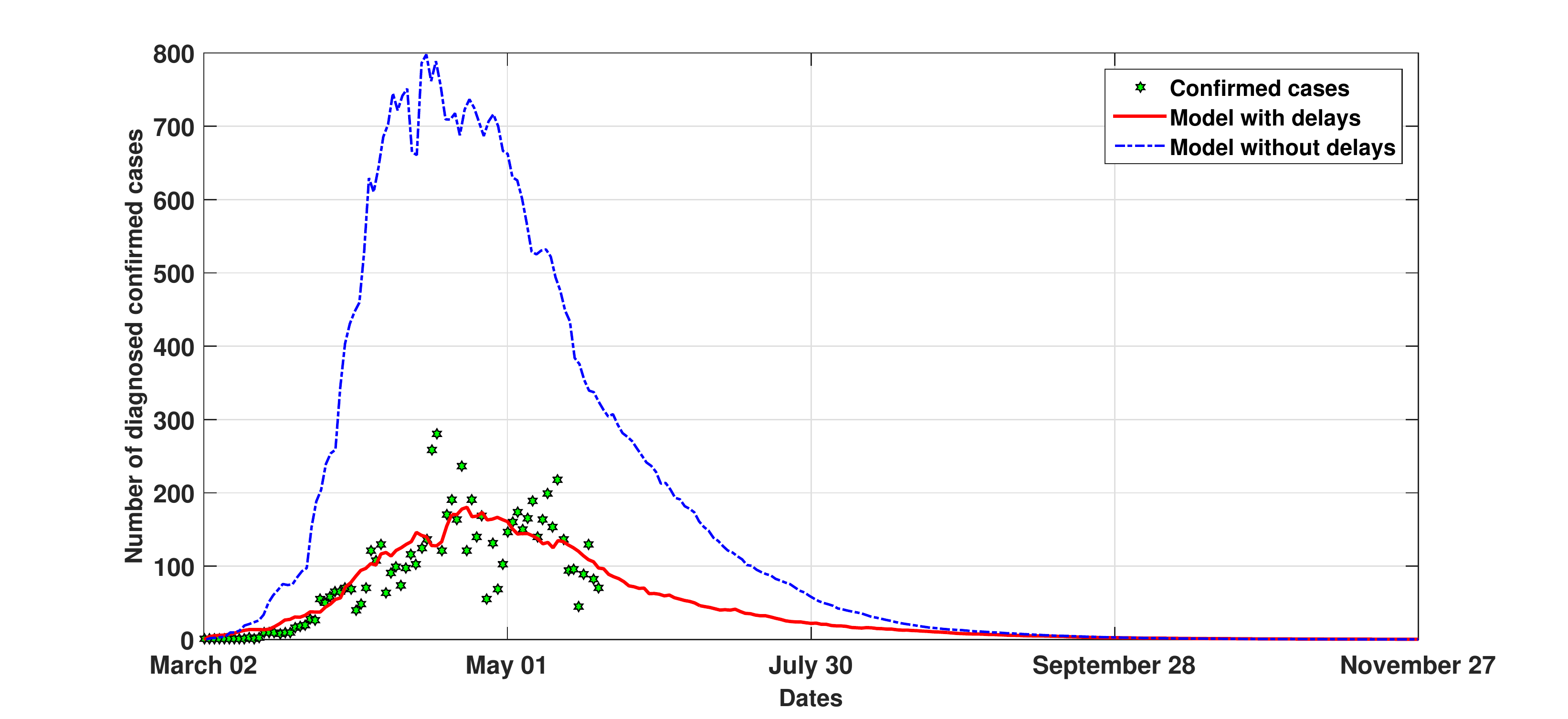}
\captionof{figure}{Effect of delays on the diagnosed confirmed cases.}
\label{Delays}
\end{figure}

To prove the biological importance of delay parameters, we give the
graphical results of Figure~\ref{Delays}, which describe the evolution
of diagnosed positive cases with and without delays.
We observe in Figure~\ref{Delays}, a high impact of delays on the number
of diagnosed positive cases, thereby the plot of model (\ref{Sys2})
without delays $(\tau_i=0,\ i=1,2,3,4$) is very different to that of the clinical data.
Thus, we conclude that delays play an important role in the study
of the dynamic behavior of COVID-19 worldwide, especially in Morocco,
and allow us to better understand the reality.

\begin{figure}[H]
\centering	
\includegraphics[width=15cm,height=7cm]{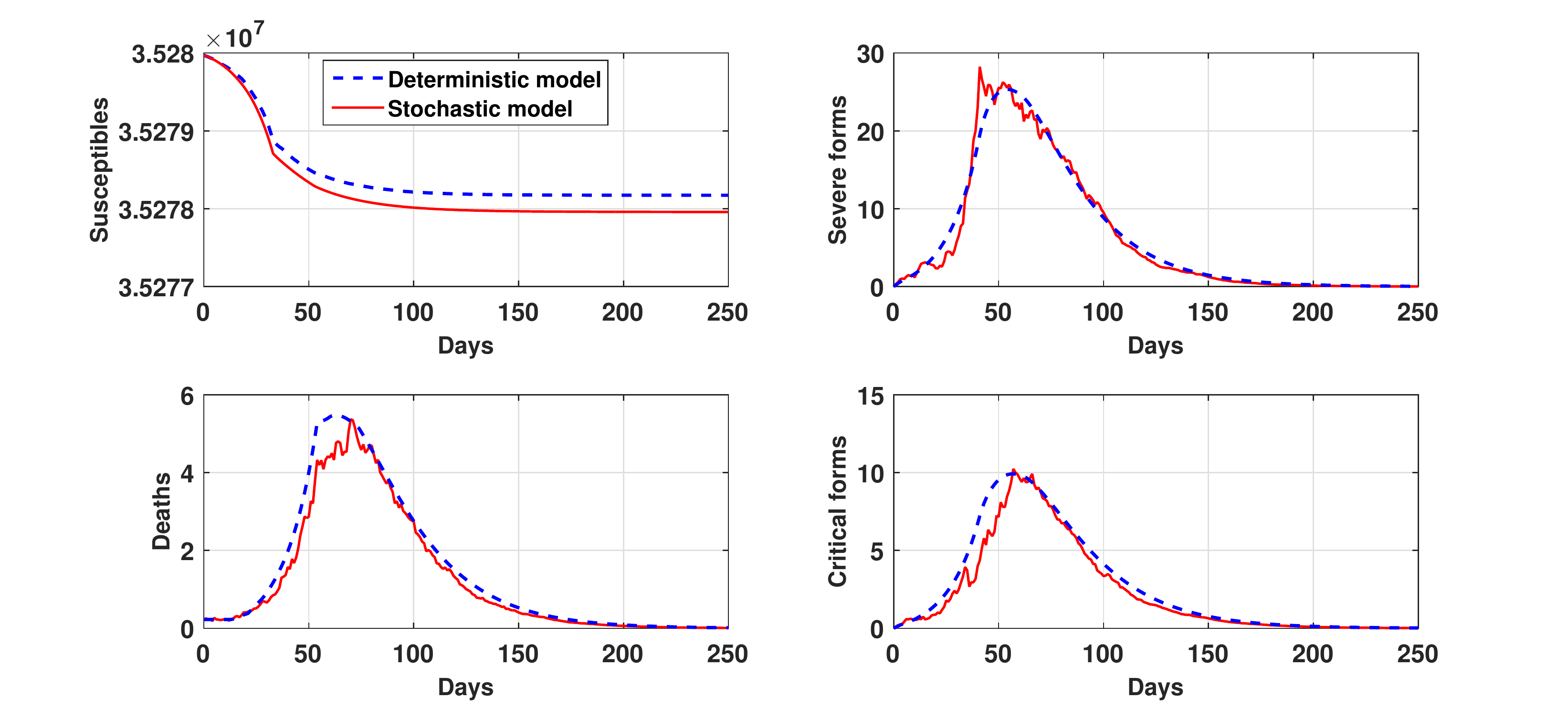}
\captionof{figure}{The evolution of susceptible, deaths,
severe and critical forms from 2 March 2020.}
\label{Evolution}
\end{figure}

In Figure~\ref{Evolution}, we present the forecast of susceptible,
severe forms of deaths and critical forms, from which we deduce
that COVID-19 will not attack the total population.
\vspace{-6pt}

In addition, the number of hospitalization beds
or artificial respiration apparatus required
can be estimated by the number of different clinical forms. 
Moreover, we see that the number of deaths given by the model 
is less than those declared in other countries \cite{Wold}, which shows that Morocco
has avoided a dramatic epidemic situation by imposing the described strategies.

\begin{figure}[H]
\centering	
\includegraphics[width=15cm,height=7cm]{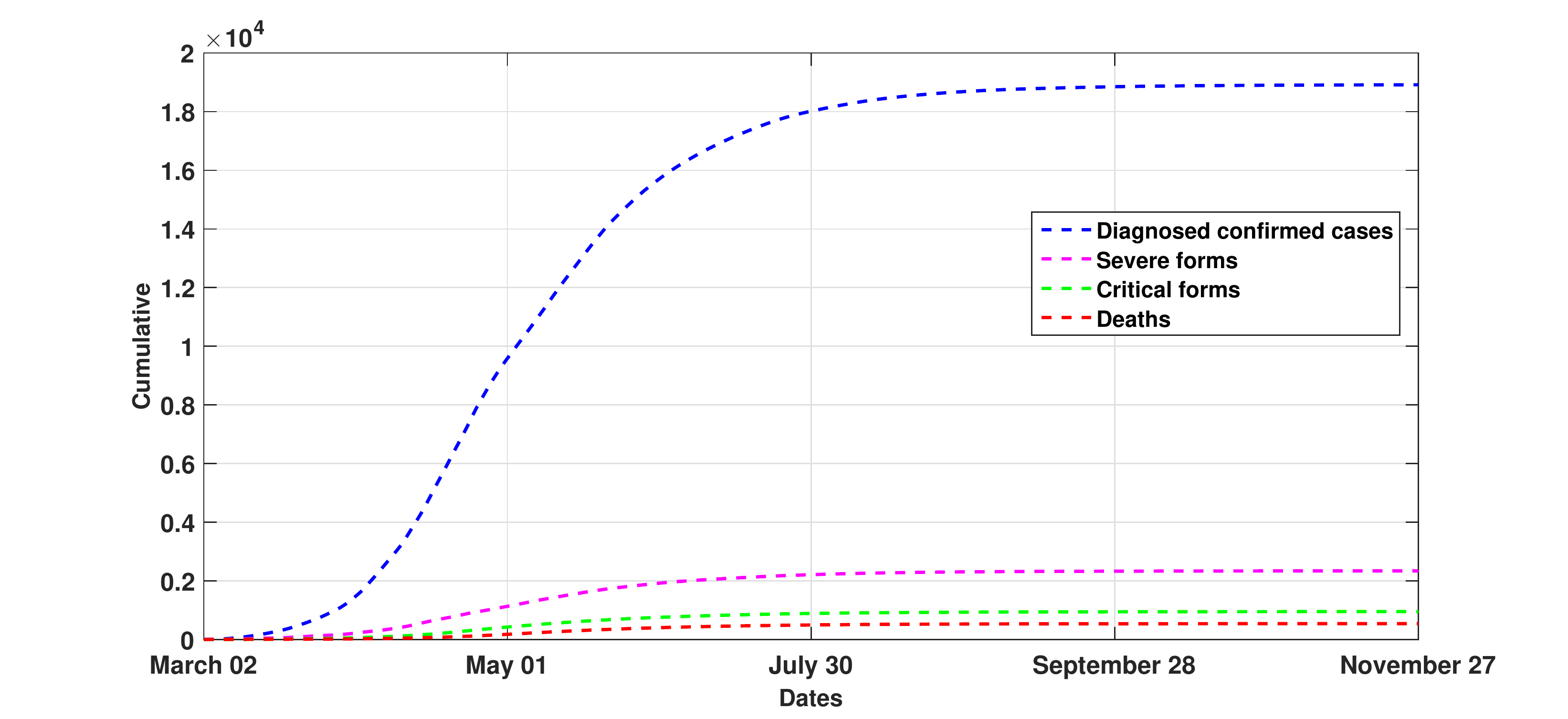}
\captionof{figure}{Cumulative diagnosed cases, severe forms,
critical forms and deaths 240 days from the start of the COVID-19
pandemic in Morocco.}
\label{Cumulative}
\end{figure}

Finally, we present in Figure~\ref{Cumulative}, the cumulative diagnosed cases,
severe forms, deaths and critical forms 240 days from the start of the pandemic
in Morocco. We summarize some important numbers in Table~\ref{CumulativeTab},
which gives us some information about the future epidemic situation in Morocco.
\vspace{-6pt}
\begin{specialtable}[H]
\centering	
\tablesize{\small}
\captionof{table}{Estimated peaks and cumulative of diagnosed cases,
severe forms, critical forms and deaths.}
\label{CumulativeTab}
\setlength{\tabcolsep}{12.6mm}
\begin{tabular}{lll}
\toprule
\textbf{Compartments} & \textbf{Peak} &  \textbf{Cumulative} \\
\midrule
Diagnosed & Around $190 $  & 18,890  \\

Severe forms & Around $ 28 $ & $ 2233 $ \\

Critical forms & Around $ 10 $  &  $ 997 $\\

Deaths & Around $ 5 $ &$ 468 $\\
\bottomrule
\end{tabular}
\end{specialtable}


\section{Conclusions}
\label{sec:6}

In this study, we proposed a new deterministic model with delay
and its corresponding stochastic model to describe the dynamic
behavior of COVID-19 in Morocco. These models provide us with
the evolution and prediction of important categories 
of individuals to be monitored, namely, 
the positive diagnosed cases, which can help to examine the efficiency
of the measures implemented in Morocco, and the different developed forms,
which can quantify the capacity of the public health system 
as well as the number of new deaths. Firstly, we have shown
that our models are mathematically and biologically well posed
by proving global existence and uniqueness of positive solutions.
Secondly, the extinction of the disease was established.
By analyzing the characteristic equation, we proved that
if $\mathcal{R}_{0} <1$, then the disease free equilibrium
of the deterministic model is locally asymptotically stable
(\mbox{Theorem~\ref{thm:3.1}}). Based on the Lyapunov analysis method,
a sufficient condition for the extinction was obtained
in the stochastic case (Theorem~\ref{thm:3.2}). Thirdly,
and since there is a substantial interest in estimating the parameters,
we applied the least square method to determine the confidence interval
of the transmission rate $\beta$ as $0.4517$ (\mbox{95\%CI, 0.4484--0.455}). In addition,
the rest of the parameters were either assumed, based on some daily observations,
or taken from the available literature. Finally, some numerical simulations
were performed to gather information in order to be able to fight against
the propagation of the new coronavirus. In 12 May 2020, the basic reproduction
number was less than one ($\mathcal{R}_{0}=0.5230$), which means that the epidemic
was tending toward eradication, which is conditional on strict compliance
with the implemented measures. Currently, the consequences of the measures
taken against COVID-19 in Morocco encourage their maintenance to control
the spread of the epidemic and quickly move towards extinction.

As future work, we intend to study the regional evolution of COVID-19 
in Morocco.


\vspace{6pt}


\authorcontributions{Conceptualization, M.M., A.B., 
H.Z., E.M.L., D.F.M.T. and N.Y.; 
Formal analysis, M.M., A.B., H.Z., 
E.M.L., D.F.M.T. and N.Y.; Investigation, 
M.M., A.B., H.Z., E.M.L., 
D.F.M.T. and N.Y.; Writing---original draft, 
M.M., A.B., H.Z., E.M.L., 
D.F.M.T. and N.Y.; Writing---review \& editing, 
M.M., A.B., H.Z., 
E.M.L., D.F.M.T. and N.Y.
All authors participated in the writing and reviewing of the paper. 
All authors have read and agreed to the published version of the manuscript.}

\funding{H.Z. and D.F.M.T. were supported by FCT within project UIDB/04106/2020 (CIDMA).}

\institutionalreview{Not applicable.}

\informedconsent{Not applicable.}


\dataavailability{The data used in this study is available from the Government of Morocco, 
being given in Figure~\ref{Historique}.}

\acknowledgments{We would like to express our gratitude to the editor 
and the anonymous reviewers, for their constructive comments and suggestions, 
which helped us to enrich the paper.}

\conflictsofinterest{The authors declare no conflict of interest.
The funders had no role in the design of the study; in the collection, analyses, 
or interpretation of data; in the writing of the manuscript, 
or in the decision to publish the results.}


\reftitle{References}


\end{document}